\documentclass[a4paper,american]{scrartcl}
\usepackage{tikz-cd}
\usepackage{verbatim}
\usepackage{lmodern}
\usepackage[utf8]{inputenc}
\usepackage[T1]{fontenc}
\usepackage{color}
\usepackage{todonotes}
\usepackage{babel}
\usepackage{amsfonts}
\usepackage{amsmath}
\usepackage{amssymb}
\usepackage{setspace}
\usepackage[authoryear]{natbib}
\usepackage[font=small]{quoting}
\usepackage{latexsym}
\usepackage[bookmarks=false,pdfborder={0 0 0},colorlinks=false]{hyperref}
\usepackage[normalem]{ulem}
\usepackage{enumerate}
\date{}

\makeatletter

\newcounter{contatore}
\newtheorem{ax}[contatore]{Axiom}

\newcounter{contatore1}
\newcounter{contatore2}
\makeatother

\begin{document}

\title{Leibnizian relationalism for general relativistic physics}

\author{Antonio Vassallo%
\thanks{Université de Lausanne, Section de philosophie,
CH-1015 Lausanne. E-mail:
\protect\href{mailto:Antonio.Vassallo@unil.ch}{Antonio.Vassallo@unil.ch}}%
, Michael Esfeld%
\thanks{Université de Lausanne, Section de philosophie,
CH-1015 Lausanne. E-mail: \protect\href{mailto:Michael-Andreas.Esfeld@unil.ch}{Michael-Andreas.Esfeld@unil.ch}}%
}

\maketitle
\begin{abstract} 

An ontology of Leibnizian relationalism, consisting in distance relations among sparse matter points and their change only, is well recognized as a serious option in the context of classical mechanics. In this paper, we investigate how this ontology fares when it comes to general relativistic physics. Using a Humean strategy, we regard the gravitational field as a means to represent the overall change in the distance relations among point particles in a way that achieves the best combination of being simple and being informative. \medskip

    \noindent \emph{Keywords}: Leibnizian relationalism, Humeanism, general relativity, gravity, shape dynamics, duality
    \end{abstract}

\tableofcontents{}

\section{Leibnizian relationalism}
\label{sec:intro}

Leibnizian relationalism is the view that there are only distance relations among sparse discrete, unextended objects -- let's call them ``matter points''.  There neither is an underlying space in which these objects and their relations are embedded nor an underlying time in which these relations evolve. Leibnizian relationalism has been considered hitherto only in the framework of classical mechanics. The aim of this paper is to investigate whether and how this view can be applied to general relativistic physics. We are not concerned with justifying Leibnizian relationalism. Whatever the merits and drawbacks of relationalism in general may be, we take it that from a systematic point of view, Leibnizian relationalism is the strongest form of relationalism and that from a historical point of view, it has been established as a serious option in the framework of classical mechanics. These facts motivate to enquire whether and how this view can be put to work in general relativistic physics as well, independently of whether or not one eventually endorses Leibnizian relationalism. 

In the following, we set out a philosophical interpretation of general relativistic physics as being committed to no more than the ontology of Leibnizian relationalism, thereby also making use of the duality between general relativity theory (GR) and shape dynamics. Our aim is to show how a non-relationalist theory like GR can go with a relationalist ontology, thus having the best of two worlds so to speak: the standard physical theory of gravitation, and a most parsimonious philosophical proposal for an ontology of the natural world. 

In order to bring out Leibnizian relationalism as the strongest form of relationism, it can be formulated in terms of only the following two axioms:

 \begin{ax}\label{a1}
    There are distance relations that individuate objects, namely matter points.
  \end{ax}

  \begin{ax}\label{a2}
     The matter points are permanent, with the distances between them changing.
  \end{ax} 

There hence is no ontological commitment to any surplus structure: by the axioms of this ontology, any ontological difference is a physical difference in the configuration of matter. On this basis, the attractiveness of Leibnizian relationalism consists in the fact that it is a very parsimonious ontology of the natural world. If there is a plurality of objects, there has to be a certain type of relations in virtue of which these objects make up a configuration that then is the world. When it comes to the natural world, the issue are relations that qualify as providing for extension. That is the reason to single out distance relations. In virtue of these relations, there is a configuration of matter that is constituted through \emph{variation} in the distances that connect sparse, unextended objects and that make it that these are matter points. \emph{Change} in these relations then is sufficient for empirical adequacy, given that all the evidence in physics comes down to relative particle positions and their change. That is why the two axioms above are minimally sufficient to formulate an ontology of the natural world.

The distance relation is irreflexive, symmetric and connex (meaning that all particles in a configuration must be related). Moreover, any representation of distances has to satisfy the triangle inequality. Any further means used to characterize these relations is just descriptive fluff, such as e.g. using the Euclidean distance formula to label the relations in a configuration. Employing the distances to individuate the objects that stand in these relations implies this: if matter point $i$ is distinct from matter point $j$, there exists at least one other matter point $k$ such that matter points $i$ and $j$ are distinguished by their relation to $k$. If there are only matter points connected by distances, all change is change in the distance relations among the permanent matter points. Time then derives from change. Following Leibniz, time is the order of succession (see notably third letter to Newton-Clarke, § 4, and fourth letter, § 41 in \citealp{Leibniz:1890aa}, pp. 363, 376). Hence, there is no time without change; but the change in the universal configuration of matter exhibits an order, and what makes this order temporal is that it is unique and has a direction, which means that by ``instants of time'' we just refer to an arbitrary monotonically increasing parametrization of a sequence of changes in a universal configuration. This then implies that the topology of time is absolute. However, there is no absolute metric of time, because there is no external measure of time: the only meaningful way to define a metric is to choose a reference subsystem within the universal configuration of matter relative to which the rate of change in distance relations is measured.          

Employing the distance relations to individuate the physical objects so that the Leibnizian principle of the identity of indiscernables is respected comes at a price: any model of this ontology has to include at least three matter points and has to comply with the mentioned requirements on the distance relation. In particular, a symmetrical universal configuration of matter is thus ruled out, including a dynamics that leads to such a configuration. However, this is no objectionable restriction: having empirical adequacy in mind, there is no need to admit, e.g., worlds with only one or two objects or entirely symmetrical worlds as physically possible worlds. This consequence just underlines that this ontology is most parsimonious, avoiding the commitment to any surplus structure. 

The ontology is exhausted by the two mentioned axioms. These axioms do not include any primitive geometrical requirements: they do not suggest, for instance, a three-dimensional or four-dimensional space, or Euclidean geometry, to represent the distances and their change. The choice of a geometry is our construction of a space to achieve a representation of the change in the distance relations in the overall configuration of matter throughout the history of the universe that is as simple and as informative about that change as possible. The same goes for dynamical parameters that are attributed to the matter points taken individually (e.g. mass, charge, spin) or to their configuration (e.g. total energy, a wave function), constants of nature, forces, fields and the like: they do not belong to the ontology. They are our means to formulate laws in the guise of dynamical equations that allow us to represent the evolution of the configuration of matter in a manner that is both as simple and as informative as possible.  

\section{Relationalist ontology and non-relationalist physical theory}
\label{sec:gr}

Leibnizian relationalism as defined by the two axioms above is an example of what a naturalized metaphysics can be like, being cut down to what is minimally sufficient to account for the evidence that we have of the natural world. However, it is opposed to a positivist metaphysics in the following sense: it rejects reading ontological commitments off from the formalisms of physical theories. The formalisms are not the guide to ontology, but parsimony and coherence together with empirical adequacy are. The reason is that simplicity in ontology and simplicity in representation pull in opposite directions: using only the concepts that describe what there is on the simplest ontology (matter points individuated by distance relations), the description of the evolution of the configuration of matter would not be simple at all, since one could only write down an extremely long list that enumerates all the change. Reading one's ontological commitments off from the simplest description -- such as e.g. Newtonian mechanics --, the ontology would not be simple at all: it would in this case be committed to absolute space and time, to momenta, gravitational masses, forces, etc.

Consequently, it is a misunderstanding to require from a relationalist ontology that it should lead to a relationalist physical theory in the sense of a physical theory that employs only the means that define the relationalist ontology. If relationalism is pushed as far as done here for it to come out as a most parsimonious ontology of the natural world that avoids any commitment to surplus structure, this is outright impossible. No given configuration of matter, as defined only by the distance relations among the matter points, contains information about the change in these relations.

Furthermore, even the most detailed relationalist physical theory, namely the best-matching dynamics worked out by \cite{Barbour:1982aa} and \cite{Barbour:2012aa}, cannot avoid the commitment to primitive geometrical facts: the relational configuration space that Barbour constructs is called ``shape space'', because each configuration in it is individuated by its form and not by its size. However, the very concept of shape requires primitive facts about angles to be meaningful, so that Barbour's ontology has to include a conformal structure. This is no criticism of Barbour's project. It just underlines that when it comes to a physical theory, more representational means are needed than what is provided by the sparse Leibnizian ontology, although this ontology is fully empirically adequate.

One can have the best of these two worlds -- i.e. a parsimonious ontology and a simple physical theory -- by putting the Leibnizian relationalist ontology in the framework of Humeanism, although we are not concerned here with the metaphysics of modality: the network of distance relations among matter points and its change throughout the entire history of the universe is the Humean mosaic. Everything else supervenes on this mosaic in the sense that it comes in as a means to represent that change in a way that achieves the best combination of being simple and being informative. Consequently, whatever geometrical and dynamical parameters a physical theory employs over and above distance relations and their change, these parameters do not belong to the ontology, but are only a descriptive means. In a nutshell, Humeanism allows us to interpret a non-relationalist physical theory in a cogent manner that is consistent with scientific realism as being committed to no more than a parsimonious relationalist ontology.

\cite{Huggett:2006aa} applied this strategy to Newtonian mechanics. The decisive step is to realize that genuinely spatiotemporal properties such as absolute acceleration and geometrical facts regarding the embedding of a relational configuration in an absolute space do not supervene on any given relational configuration, but on the \emph{entire} history of relational change. This history exhibits certain patterns or regularities, which enable us to single out the idea of inertial motion in a background space as a particularly simple and regular motion. \cite{Huggett:2006aa} employs the concept of adapted frame, that is, a reference frame tied to a body (e.g. a material particle). More precisely, an adapted frame amounts to an assignment of real numbers at a given time, that is, a set of $N$-tuples such that (i) the origin of the frame -- the $(0,0,0,\dots,0)$ tuple -- corresponds to the ``position'' at time zero of the body the frame is adapted to and (ii) the distances along the $N$ axes correspond to the distances (i.e. the numerical labeling of distance relations) from the body. Inertial frames then are those frames in which an observer would describe the history of relations in the simplest and strongest way in terms of Newton's laws.  Note that there might be no body tied to an inertial frame; nonetheless, any other adapted frame can be related to an inertial one by means of a spatial rigid translation. Roughly, a spatial translation from a frame $O$ into another one $O'$ amounts to ``shifting'' the set of $N$-tuples constituting $O$ by a certain factor (defined, e.g., by a suitable continuous function that takes an $N$-tuple in $O$ as input and gives back an $N$-tuple in $O'$), such that the relative distances between particles remain unaltered in the new frame.

It is important to stress once again the fact that this construction does not require any ontological commitment over and above those entailed by axioms 1 and 2. Choices such as the dimensionality $N$ of space or the numerical labeling of distance relations in a configuration are in fact entirely arbitrary, the only constraint being, obviously, that such choices enable the best possible physical description. Similarly, the definition of spatial rigid translation relies only on arbitrary numerical assignments and, hence, does not require any pre-existing geometrical notion such as that of affine connection.

Once this characterization of inertial frame is in place, absolute acceleration can be reduced to the history of change of the spatial relations holding between an inertial and a non-inertial frame: acceleration is part of the description of how much the pattern that constitutes the history of a non-inertial frame deviates from the regularities encoded in the inertial pattern as seen in an inertial frame. By the same token, any other absolute quantity of motion such as, for example, angular momentum, can be shown to supervene on the Humean mosaic.

Similarly, the regularities in the history of relations make it that Euclidean geometry is the simplest and most informative geometry representing an embedding of that history in an external space. This reasoning can be applied not only to space-like geometrical facts, but also to time-like ones, as indicated in the preceding section: the history of change in the distance relations in the universal configuration of matter exhibits an order that is unique and directed. Time derives from this order in the sense that temporal facts supervene on the history, i.e. an ordered sequence of instantaneous distance relations making up the universal configuration of matter.

The same Humean strategy can be applied to dynamical parameters such as mass and charge: these are means of representing the change in the spatial relations in the system that achieves the best combination of being simple and being informative about that change (see \citealp{Hall:2009aa}, § 5.2). That is to say: it is not mass and charge qua properties belonging to individual matter points that determine their trajectories by means of the causal role that they play in the laws of classical mechanics and electrodynamics; on the contrary, the trajectories that the matter points trace out through the whole history of the universe make it that parameters such as mass and charge figure in the dynamical laws such that a value of mass and charge applies to the particles taken individually. Again, there is no commitment to surplus structure here in the sense that there are no situations possible in which mass and charge exist, being properties belonging to individual matter points, but do not manifest themselves in any change in the distance relations among the matter points. Mass and charge, etc. only are our means of representation of the change that actually occurs.

Furthermore, this Humean strategy has recently been applied to quantum mechanics, namely to the quantum theories that solve the measurement problem by being committed to a primitive ontology of matter distributed in physical space (such as particle positions in Bohmian mechanics): that distribution and its evolution then is the Humean mosaic. The -- universal -- wave function supervenes on that mosaic, being a means to achieve a description of that evolution that is both most simple and most informative, but does not belong to the ontology (see \citealp{Miller:2014aa}, \citealp{Esfeld:2014aa}, \citealp{Callender:2014aa}, and \citealp{Bhogal:2015aa}).

Again, our aim here is not to defend Humeanism. We only take up the fact that the outlined sparse Humean strategy is by now well established as an option for classical and quantum mechanics in the literature, whatever the pros and cons of this strategy may eventually be. We employ this strategy here because we consider it to be, as things stand, the only promising means to vindicate Leibnizian relationalism for general relativistic physics, since this strategy shows a clear-cut way how to philosophically interpret a non-relationalist physical theory as being committed to no more than a sparse relationalist ontology. In particular, the sparse Humeanism that takes the Humean mosaic to consist only in distance relations among matter points and the change of these relations seems to be, as things stand, the only strategy that enables us to do without a commitment to primitive geometrical facts in the ontology and to dismiss fields, including notably the gravitational or metrical field of GR, from the ontology, regarding them as representational means that supervene on the overall history of the change in the distance relations among the matter points.

\section{Leibnizian relationalism for general relativity: the challenges}
\label{sec:challenges}

The metric formulation of GR involves a set of tensor-fields defined over a four-dimensional semi-Riemannian manifold. Using a local (i.e. component-based) language, we can cast the field equations of the theory in the following form ($\kappa$ being an appropriate constant):
\begin{equation}\label{eq:efe}
G_{\mu\nu}[g_{\mu\nu}]=\kappa T_{\mu\nu}[\phi, g_{\mu\nu}].
\end{equation}
For simplicity's sake, we omit the term involving the cosmological constant. The indices $\mu$ and $\nu$ range from $0$ to $3$. The left hand side of \eqref{eq:efe}, being dependent on the metric tensor $g_{\mu\nu}$, contains information about the geometry of the manifold, while the right hand side conveys information about the physical properties (e.g. energy density, momentum flux) of the material sources mathematically modeled by the portmanteau $\phi$, such as the electromagnetic field.

This formulation makes clear that Leibnizian relationalism combined with the Humean strategy faces what appears to be two knock down objections: in the first place, \emph{spacetime in GR is inherently four-dimensional}. There is no objective -- that is, non-arbitrary -- way to distinguish spatial from temporal relations holding between events. This in turn implies that there is no well-defined notion of spatial configuration in a Leibnizian sense, which undermines the possibility to recover time in the way \cite{Huggett:2006aa} does for Newtonian mechanics. Moreover, \emph{GR is considered as the paradigm example of a field theory}. Accordingly, insofar as there is a relationalism worked out for GR in the literature, it is a relationalism that replaces spatial with spatiotemporal relations and particles with fields (see \citealp{Rovelli:1997aa}, for a prominent example and \citealp{Pooley:2001aa}, and \citealp{Pooley:2013aa}, section 7, for a philosophical assessment).

Following this field relationalism, there is a plenum of fields instead of distance relations between sparse points. However, one may wonder whether this relationalism is worth its name, because there no longer is a principled difference with substantivalism. The reason is the ambiguous status of the metrical field in GR: the metrical field implements the geometry of spacetime, but it also carries energy and momentum. For the substantivalist, the metrical field has a special status, being spacetime, since it contains the geometry of the universe. For the field relationalist, it is a field interacting with other fields. Both endorse the metrical field as an entity \emph{sui generis}, the substantivalist calling it ``spacetime'', the field relationalist regarding it as a field among other fields.


The Leibnizian can resist the move to such a field relationalism: in GR as in any other field theory, fields are tested by the motion of particles. There is no direct evidence of fields. All the evidence is one of particle motion. Thus, all empirical determinations of the gravitational field amount to observations of the motion of bodies in the sense of change in the instantaneous spatial relations they stand in. As Einstein puts it, \begin{quote} The gravitational field manifests itself in the motion of bodies. Therefore the problem of determining the motion of such bodies from the field equations alone is of fundamental importance. (\citealp{Einstein:1949aa}, p. 209)\end{quote} 

Consequently, the field equations are there to determine the motion of bodies. This opens the door for applying the Humean strategy also to GR: instead of buying into the dualism of gravitational field and material bodies as suggested by Einstein and Infeld in this quotation, one can maintain that the ontological bedrock -- the Humean mosaic -- consists in the motion of bodies only. The gravitational field is a mere representational means that enables us to describe the overall motion of the bodies in a manner that optimizes simplicity and information about that motion. Hence, the objection against a particle ontology from GR being a field theory takes a piece of description (fields) for a piece of ontology. This reasoning also applies to the account of purely gravitational phenomena, such as spacetime singularities or gravitational waves. Also in these cases all that is physically observable is the change of spatial relations among material bodies.

As mentioned in section~\ref{sec:intro}, this Leibnizian stance cannot recognize all mathematically possible solutions of in this case the field equations of GR as describing physically possible worlds. In particular, \eqref{eq:efe} allows for empty cosmological solutions, that is, models in which the universe is totally deprived of matter, yet spacetime has, say, ripples and lumps. These solutions have to be dismissed as mathematical surplus structure of the theory; taking them ontologically serious would amount to inflating the ontology with the gravitational field as a substance \emph{sui generis} existing over and above matter. Again, dismissing these solutions is no problem, since the world we live in undoubtedly is not an empty universe. By the same token, also the solutions lacking (global) Cauchy hypersurfaces -- that is, $3$-surfaces that are intersected only once by any non-space-like curve -- have to be dismissed, since these solutions would depict cosmological models in which it is even not possible to give a physical meaning to a space/time split.


Although all these solutions can be rejected as mathematical surplus without ontological significance and although the objection from GR being a field theory can be countered on the basis of all empirical evidence consisting in the motion of bodies, the fact remains that the fundamental relations in GR seem to be spatiotemporal ones between events rather than distance relations between particles that change. This comes out clearly even  if one focuses only on models of spacetimes that admit space-like Cauchy surfaces. Consider a solution of \eqref{eq:efe} consisting in a $4$-geometry that includes wordlines of material bodies, such that the manifold possesses a product topology $\Sigma_{3}\times\mathbb{R}$. Then, we can reduce the physical description of this $4$-geometry to the sum of descriptions on a pile of three-dimensional Cauchy surfaces $\Sigma_{3}$ that cut the manifold in the space-like direction. The manifold can thus be foliated by a series of space-like leaves. Nonetheless, the inherent four-dimensionality of GR shows up in the fact that the choice of the foliation is not unique. In fact, for any sequence of $\Sigma_{3}$ that we select, we always recover the same four-dimensional physical representation including the worldlines of the material bodies.

Consequently, the following result seems to impose itself upon us: instead of matter points that are each separated by a distance from each other, with that distance individuating the matter points and changing, there are continuous sequences of matter point events, forming continuous lines (worldlines) that are individuated by the spatiotemporal intervals between them. These intervals do not change. They all exist at once. That is why this ontology is known as \emph{block universe}: all the events throughout the history of the universe exist at once. This four-dimensional ontology can be cast in terms of a spacetime relationalism. No commitment to a spacetime that exists in distinction from matter, with matter filling spacetime, is called for. Indeed, the famous hole argument, going back in its contemporary form to \cite{Earman:1987aa}, can be seen as making a case for such a relationalism.

That notwithstanding, the relationalism endorsed by the proponents of the hole argument relies on a huge amount of geometrical structure, as does the field relationalism mentioned above: these sorts of relationalism are committed to the metrical field being part of the ontology, even if that field is considered as consisting in metrical relations between point-events. By contrast, the Leibnizian relationalism advocated here does not countenance any primitive geometrical facts. It accepts distance relations and their change as primitive, whereas the geometry is a mere means of representing that change: the geometry is fixed together with the dynamics as the means to achieve a description of that change that strikes the best balance between being simple and being informative. Thus, distances neither are inherently three-dimensional or four-dimensional, nor are they tied to Euclidean or Riemannian geometry. This ontology is centred only on the distinction between distance relations, providing for variation within the configuration of matter by individuating the matter points, and change of these relations. That distinction is crucial for Leibnizian relationalism to achieve empirical adequacy.

By contrast, the distinction between variation within a given configuration and change of that configuration gets lost in the block universe ontology. If one replaces point particles with point events forming continuous lines, and a $3$-geometry with a $4$-geometry to represent the relations between them, one obtains variation, but no change: the relations represented in terms of a $4$-geometry provide for variation within the block universe. However, since these relations exist all at once, there is no change. Only when we cut three-dimensional slices through the block and compare them, we can define change in terms of the differences between such slices. But this change concerns only an -- arbitrary -- description and not the ontology, since the ontology is the four-dimensional block.

To include change, one therefore has to stipulate the following over and above the global geometrical order characterizing the block universe (whatever the number of dimensions may be in which that geometry is formulated): the points on each worldline are ordered according to earlier and later, with that order being unique and directed. Then there is a local time for each object (worldline), but no global time. Comparing different points on each worldline, one can introduce change in terms of different spatiotemporal intervals between different points on two worldlines. In this way, the spatiotemporal relations provide for variation and change at once against the background of the points on each worldline being ordered according to earlier and later.

There are, however, at least two drawbacks of this move: (a) temporal order has to be presupposed as a primitive in this case, that is, the ordering of the points on a worldline according to earlier and later. That ordering obtains as a primitive matter of fact, independently of whether or not there is change in the spatiotemporal intervals between these two lines. Hence, temporal order is not derived from change, but change is derived from the order of the points on each worldline being temporal. What is the difference between ordering the points on such a line according to, say, below and above and ordering them according to earlier and later? That difference is primitive. Consequently, one cannot do without endorsing a primitive temporal, metrical requirement on the block universe ontology, even if one casts this ontology in relationalist terms instead of endorsing a spacetime substance.

(b) The question then is whether endorsing a primitive temporal order of the points on each worldline is sufficient for empirical adequacy. There is no possibility to introduce becoming and the passage of time: there is a local temporal order of the points on each worldline, but since these points exist all at once, there is no such thing as becoming in the ontrology. Becoming and the passage of time have to be relegated to consciousness being a point moving on such a line in a fixed direction. As Weyl famously put it, \begin{quote} The objective world simply \emph{is}, it does not \emph{happen}. Only to the gaze of my consciousness, crawling upward along the life line of my body, does a section of this world come to life as a fleeting image in space which continuously changes in time (\citealp{Weyl:1949aa}, p. 116).\end{quote}

Consequently, it is doubtful whether paying the price of a primitive temporal order is worthwhile, since doing so does not give us the phenomenology of time. Leibnizian relationalism, by contrast, easily recovers this phenomenology on the basis of endorsing change as primitive.

These considerations are by no means conclusive, but they show again that the ontology of the natural world is a matter of philosophical argument and cannot be read off from the formalism of a physical theory. When it comes to ontology, it is worthwhile investigating the option to take the $4$-geometry as a means of representation instead of endorsing it as constituting the ontology of GR in the guise of a block universe. It is true that the distances making up the Leibnizian configuration of matter points and their change cannot be specified uniquely in GR. However, this may just show that fundamental space-like facts can be \emph{described} in different yet equivalently simple and strong ways. One should not conclude from this descriptive underdetermination that there are no fundamental space-like, Leibnizian facts, on pain of losing the parsimony of the ontology.

This situation is in a certain sense analogous to the one of the quantum Humeanism mentioned at the end of the preceding section: on Humeanism applied to Bohmian mechanics, for instance, the particle positions and their change throughout the history of the universe make up the Humean mosaic. However, due to the universe being in quantum equilibrium, our knowledge is limited to what can be obtained by applying Born's rule. Furthermore, there are at least two different formulations of Bohmian mechanics -- the standard one and the identity-based one -- that agree on the accessible facts, but disagree on the particle trajectories (see \citealp{Goldstein:2005b, Goldstein:2005a}). Thus, the Humean mosaic cannot be uniquely specified in the Bohmian primitive ontology approach. In general, a limit of accessibility applies to any primitive ontology proposed for quantum mechanics (see \citealp{Cowan:2015aa}). Nonetheless, the argument for these theories is that they solve the measurement problem by providing an ontology and a dynamics that make unique measurement outcomes intelligible. Hence, the issue of what is the Humean mosaic is one of ontological argument and not whether or not the Humean mosaic can be uniquely described or is accessible to observation. 

Indeed, our main access to the Leibnizian relations is through the way in which they change, that is, through the evolution of the distance relations among the matter points. It may then turn out that the simplest and most informative description of that evolution is formulated in terms of instantaneous action at a distance, thus involving a unique specification of the spatial relations among the particles, as in classical gravitation; or it may turn out that the simplest and most informative description of that evolution is formulated in terms of retarded action, as in field theories from classical electromagnetism on (or in terms of retarded and advanced action, as in relativistic theories of direct particle interaction). In this case, the instantaneous distance relations among the particles do not figure in the dynamics, although it is their change that the dynamics seeks to capture. Again, this is a matter of what type of description turns out to achieve the best balance between being simple and being informative about the change in the empirical world, instead of being a matter of ontology. The measure for ontology is parsimony and coherence together with empirical adequacy, not the dynamical variables that figure in a particular physical theory.

\section{Interlude: shape dynamics}
\label{sec:shape}

Recent progress in physics provides a further argument to vindicate the Leibnizian perspective on space and time in general relativistic physics. \citet{Gomes:2011aa} show that there is an alternative theory defined on the phase space of GR. Such a theory is not just alternative, but \emph{dual} to GR in the sense that, under the appropriate choice of gauge, the dynamical trajectories of the two theories coincide. This dual theory is Barbour's shape dynamics (see \citealp{Gomes:2013aa}, for a concise presentation).

Leaving aside technical considerations, the important philosophical consequence of this duality in classical (that is, non-quantum) gravity is that there are two metaphysical stances with respect to space and time that are compatible with the empirical predictions of general relativistic physics. The first one is the irreducibly four-dimensional perspective of GR, the second one is the three-dimensional perspective of shape dynamics. In this theory, dynamics can be depicted as a succession of (conformal) $3$-geometries. Unlike GR, this theory is \emph{not} invariant under change of foliation. Instead, this characteristic symmetry of GR is traded for a local conformal symmetry, which means that the geometry defined on each folio lacks a privileged notion of scale -- it is not \emph{size}, but \emph{shape} that matters.

In shape dynamics, there is a well-defined notion of instantaneous spatial configuration. This fact does not imply that instants are referred to an external clock ticking in the background. What exists is a succession of spatial configurations that can be arbitrarily labeled by a monotonically increasing parameter; but this succession is not \emph{in time}. On the contrary, it defines a ``time-like'' direction for the unfolding of the dynamics. Hence, shape dynamics is compatible with Leibniz' (and Mach's) view of space as the order of coexistence (i.e. bodies related by spatial relations) and time as a bookkeeping device to describe the order of change (i.e. the succession of instantaneous spatial configurations).


We do not claim that shape dynamics fully implements the ontology of Leibnizian relationalism. First of all, also this theory is a field theory. Secondly, at this stage, the theory has been worked out only for pure gravity (but see \citealp{Gomes:2012aa}, for some first results for gravity/matter coupling). Thirdly, this theory includes irreducible geometric facts regarding angles (that is, a conformal structure) contra a fundamentally geometry-less ontology. However, as argued in section \ref{sec:gr}, it would be a misunderstanding to require that a relationalist ontology has to be vindicated by a physical theory that employs only relationalist representational means. The point at stake here is that the GR/shape dynamics duality hints at the fact that there exists a privileged way to conceive the Humean mosaic -- which corresponds to the appropriate gauge where the two theories coincide in phase space -- and that such a conception is fully compatible with a Leibnizian take on space and, most importantly, time. To put it in the terms of \citet{Gryb:2015aa}, gravity is ``Janus-faced'': it is compatible both with a block universe where there is variation of local matters of fact but no change (and, hence, no time ordering change), and with a dynamical universe where there is change in the guise of change in the fundamental relations that connect the fundamental objects, on which temporal facts supervene in the form of a monotonically increasing parameter labelling the succession. In short, the duality of GR and shape dynamics again shows that one cannot read the ontology off from the formalism of a physical theory, since one would in this case end up with two proposals for the ontology of general relativistic physics that contradict each other, based on two different formalisms for general relativistic physics. The ontology has to be settled by criteria such as parsimony and coherence together with empirical adequacy.

\section{The Humean strategy applied to general relativity}
\label{sec:Humean}

Against this background, let us now try out for GR the Humean strategy that \cite{Huggett:2006aa} developed to vindicate a Leibnizian relationalist ontology for Newtonian mechanics. The distance relations among point particles and their change throughout the entire history of the universe are the Humean mosaic. As in the classical case, we can define an adapted frame of reference as one in which a certain particle is always at rest at the origin. We can then relate all such reference frames by means of rigid spatial translations. We look as usual for regularities in the history that admit a particularly simple yet strongly informative description.

The problem that we face in this respect is that there is no such thing as classical inertial motion in GR: this theory merges inertial structure and the gravitational field. Consequently, there is no unique way to separate them, let alone find a case where the gravitational part of the sum vanishes in a region bigger than a point. Hence, the description that we search for cannot be encoded in a law of the form $\ddot{x}^\mu=0$ (a dot indicates the usual differentiation with respect to an arbitrary time-like parameter defined along a given pattern); in this case, we would be dealing with a Newtonian world rather than a general relativistic one. Nonetheless, even if we cannot make use of the concept of inertial motion in GR, still we can employ a generalization of this concept: the simplest and most informative description of the regularities in a general relativistic world is that of geodesic motion (more precisely, non-space-like geodesic motion), that is, $\ddot{x}^\mu+\Gamma^{\mu}_{\phantom{\mu}\nu\sigma}\dot{x}^\nu\dot{x}^\sigma=0$, using an appropriate (affine) parametrization. Geodesic motion is a well-defined concept in GR, since the theory stipulates that the motion of freely-falling bodies follows geodesic trajectories. Here we immediately see that the (Levi-Civita) connection with coefficients $\Gamma^{\mu}_{\phantom{\mu}\nu\sigma}$ is not a geometrical feature crafted so to speak in the Humean mosaic: it is just a tool that helps to formalize how the description will change among different observers moving geodesically -- in other words, a description of how different frames adapted to freely-falling bodies and related by rigid spatial translations -- differ.

Note that we do not presuppose the existence of a spacetime structure (in particular, a connection) that \emph{defines} what it is for a motion to be geodesic, but, rather, the other way round: we define geodesic motion as a particularly simple pattern in the entire history of relational change. Then we construct a connection as a bookkeeping device that accounts for the relational differences among different geodesics: there are no primitive geometrical facts, just descriptive tools taken from the language of (differential) geometry. Similarly to the classical case, also here there might actually be no body moving geodesically: what we require is just that any other adapted frame can be related to a geodesic one by means of a spatial rigid translation. If this is not possible, then the possible world we are dealing with does not admit the laws of general relativistic physics as theorems of the simplest and strongest system.

It is also worth noting that, in the Newtonian case, the coefficients of the connection do not arise, because all inertial motions are identical, which is consistent with the fact that Newtonian spacetime admits a flat connection (i.e. the simplest and strongest description of inertial motion is one for which $\Gamma^{\mu}_{\phantom{\mu}\nu\sigma}=0$). In the general relativistic case, by contrast, geodesic motions might in fact differ even when adopting the same affine parametrization (what is usually called ``geodesic deviation'', which is at the root of the explanation of how tidal forces arise). From this point on, all the relevant geometrical and dynamical features of the theory can be shown to supervene on the Humean mosaic in the usual manner. For example, curvature is a simple and informative way to describe relational change between freely-falling bodies. More generally, the metrical structure $g_{\mu\nu}$ of spacetime supervenes on geodesic motions through the usual relation $\Gamma^{\mu}_{\phantom{\mu}\nu\sigma}=\frac{1}
{2}g^{\gamma\mu} \big(g_{\sigma\gamma,\nu} + g_{\gamma\nu,\sigma} - g_{\nu\sigma,\gamma}\big)$ (a comma indicates standard differentiation with respect to the subsequent index). Geodesic trajectories also suffice to fix the topology of spacetime, as shown in \citet{Malament:1977aa}. Note that, by construction, the geometrical description encoded in $\Gamma^{\mu}_{\phantom{\mu}\nu\sigma}$ and $g_{\mu\nu}$ does not inhere within \emph{single} trajectories, but supervenes on the \emph{totality} of geodesic motions. In this sense, the continuous large-scale geometry of spacetime results from a coarse-grained description of the entire history of relational change among particles.

In general, all fields -- including, as we have seen, the metrical one -- are in this framework shortcuts to describe certain forms of particle motion. Hence, the relationship between particles and fields is not physical -- particles do not generate fields, nor are they pushed by fields --, but descriptive: a particle-vocabulary can be coarse grained to a field one in order to simplify the description of gravitational phenomena without loss of physical information. Clarifying this point is very important because if we took particles and fields on a par and, in particular, regarded particles as field sources, then singularities would arise. This is all the more true for GR. In this theory, a Schwarzschild radius is associated to any extended body, which depends on the body's mass: roughly speaking, if the body's extension drops below this radius, then the body collapses into a black hole singularity. It is then obvious why, in GR, this notion of massive point-particle makes little sense besides some particular approximations. However, it should be clear by now that our notion of particle is \emph{not} the one just discussed.

By the same token, quantities such as momentum densities, energy flows and the like can be constructed out of the field-vocabulary. For example, the stress-energy tensor $T_{\mu\nu}$ figuring in \eqref{eq:efe} does not refer directly to the underlying particle motions, but expresses their field-like description; it is, so to speak, a description of how a description changes. What has been said so far shows that there is nothing strange in claiming that a large-scale continuum theory such as GR (i.e. a theory used to model continuous material systems at astrophysical or cosmological scales) admits an extremely sparse ontology.


That notwithstanding, these results do not show that the central law of GR, namely the field equations \eqref{eq:efe}, can be recovered from the Humean mosaic as a theorem of the system that strikes the best balance between simplicity and strength in describing the change in the distance relations among the point particles that occurs throughout the history of the actual world. Indeed, thus recovering the field equations \eqref{eq:efe} of GR is outright impossible for the following reason: if the Humean mosaic are distance relations among sparse matter points and their change throughout the history of the universe, we can obtain geometry and dynamics in a package as the system that optimizes simplicity and information in the description of that change. However, geometry and dynamics remain distinct, as in the classical case. In other words, the dynamical laws that describe such a Humean mosaic are those of a theory formulated over a fixed (albeit non-Euclidean) background. There is nothing in the regularities of motion of such a Humean mosaic that could lead to a higher-order description where geometry is \emph{coupled} to matter in a dynamical way. The ontology of there being only distance relations and their change is too meagre to get \eqref{eq:efe} from such a possible world by Humean means: the most we can get is a possibly very complex, but still non-dynamical geometry compatible with the geodesic law of motion. This is because geometry supervenes on the mosaic by means of the law of geodesic motion, while  non-gravitational fields come out as a way to describe non-geodesic motion: given this construction, there is no way to conjecture any inter-dependence between geometry and material fields.


Nonetheless, this result can be accommodated in an ontology of there being only distance relations among sparse matter points and their change that does justice to general relativistic physics. Consider the class of cosmological solutions of \eqref{eq:efe} (modulo those that we have previously discarded as mathematical surplus): each of these models describes a possible world in which GR holds. These models are each characterized by a certain metric $g_{\mu\nu}$ (compatible with the Levi-Civita connection) and a certain stress-energy tensor $T_{\mu\nu}$. The main difference that this cluster of possible worlds bears with respect to a cluster of Newtonian worlds is that the Newtonian worlds \emph{always} feature the same spatial and temporal metrical structures. By contrast, in general, $g_{\mu\nu}$ co-varies with $T_{\mu\nu}$ from world to world in the GR case. This is one possible sense in which we can understand what is known as the background independence of GR: the same spatiotemporal structures of Newtonian mechanics appear in all Newtonian models, so that they are nomologically necessary; spatiotemporal structures in GR, by contrast, are nomologically contingent.

Against this background, what \eqref{eq:efe} represents is the manner in which in any possible world of GR, spatiotemporal structures described by some metric tensor $g_{\mu\nu}$ are correlated with material structures described by a stress-energy tensor $T_{\mu\nu}$. That is to say: in any possible world of the Leibnizian ontology of distance relations among sparse matter points and their change in which GR is valid, the system that strikes the best balance between simple and being informative about a particular such world is one in which a dynamics describes the evolution of the configuration of matter over a fixed background -- that is, a particular cosmological solution of \eqref{eq:efe} describing a particular possible world of GR. The general law \eqref{eq:efe} then is not a theorem that generalizes the regularities within a world, but a \emph{trans-world} generalization: it expresses the relationship between the mathematical structures formulating the geometry and the mathematical structures formulating the dynamics in any possible world where GR is the best system describing the change in the distance relations among the point particles. In sum, we propose a generalization of the Humean account that \cite{Huggett:2006aa} developed for Newtonian mechanics that shows that geometry and dynamics supervene on the Humean mosaic of Leibnizian relations \emph{at each world} by means of a geodesic law; we then obtain Einstein's field equations as the simplest and strongest description of the correlation between geometry and dynamics for the whole class of these worlds. We submit that this, then, is the way in which Leibnizian relationalism can be put to work in general relativistic physics, leaving open whether taking everything into account this is as serious an option for general relativistic physics as Leibnizian relationalism is for classical mechanics. 

\paragraph{Acknowledgements.} We are grateful to Dirk-André Deckert and Dustin Lazarovici for helpful discussions. A. Vassallo's work on this paper was supported by the Swiss National Science Foundation, grant no. 105212\_149650.

\bibliographystyle{apalike}
\bibliography{references_fundont}

\begin{thebibliography}{}

\bibitem[Barbour, 2012]{Barbour:2012aa}
Barbour, J. (2012).
\newblock Shape dynamics. {A}n introduction.
\newblock In Finster, F., M{\"u}ller, O., Nardmann, M., Tolksdorf, J., and
  Zeidler, E., editors, {\em Quantum field theory and gravity}, pages 257--297.
  Basel: Birkh{\"a}user.

\bibitem[Barbour and Bertotti, 1982]{Barbour:1982aa}
Barbour, J. and Bertotti, B. (1982).
\newblock Mach's principle and the structure of dynamical theories.
\newblock {\em Proceedings of the Royal Society A}, 382:295--306.

\bibitem[Bhogal and Perry, 2016]{Bhogal:2015aa}
Bhogal, H. and Perry, Z.~R. (2016).
\newblock What the {H}umean should say about entanglement.
\newblock {\em No{\^u}s}.
\newblock DOI 10.1111/nous.12095.

\bibitem[Callender, 2015]{Callender:2014aa}
Callender, C. (2015).
\newblock One world, one beable.
\newblock {\em Synthese}, 192(10):3153--3177.

\bibitem[Cowan and Tumulka, 2016]{Cowan:2015aa}
Cowan, C.~W. and Tumulka, R. (2016).
\newblock Epistemology of wave function collapse in quantum physics.
\newblock {\em British Journal for the Philosophy of Science}, 67:405--434.

\bibitem[Earman and Norton, 1987]{Earman:1987aa}
Earman, J. and Norton, J. (1987).
\newblock What price spacetime substantivalism? {T}he hole story.
\newblock {\em British Journal for the Philosophy of Science}, 38:515--525.

\bibitem[Einstein and Infeld, 1949]{Einstein:1949aa}
Einstein, A. and Infeld, L. (1949).
\newblock On the motion of particles in general relativity theory.
\newblock {\em Canadian Journal of Mathematics}, 1:209--241.

\bibitem[Esfeld, 2014]{Esfeld:2014aa}
Esfeld, M. (2014).
\newblock Quantum {H}umeanism, or: physicalism without properties.
\newblock {\em The Philosophical Quarterly}, 64(256):453--470.

\bibitem[Gerhardt, 1890]{Leibniz:1890aa}
Gerhardt, C.~I., editor (1890).
\newblock {\em Die philosophischen {S}chriften von {G}. {W}. {L}eibniz. {B}and
  7}.
\newblock Berlin: Weidmannsche Verlagsbuchhandlung.

\bibitem[Goldstein et~al., 2005a]{Goldstein:2005b}
Goldstein, S., Taylor, J., Tumulka, R., and Zangh{\`\i}, N. (2005a).
\newblock Are all particles identical?
\newblock {\em Journal of Physics A: Mathematical and General},
  38(7):1567--1576.

\bibitem[Goldstein et~al., 2005b]{Goldstein:2005a}
Goldstein, S., Taylor, J., Tumulka, R., and Zangh{\`\i}, N. (2005b).
\newblock Are all particles real?
\newblock {\em Studies in History and Philosophy of Modern Physics},
  36(1):103--112.

\bibitem[Gomes et~al., 2011]{Gomes:2011aa}
Gomes, H., Gryb, S., and Koslowski, T. (2011).
\newblock Einstein gravity as a 3{D} conformally invariant theory.
\newblock {\em Classical and Quantum Gravity}, 28:045005.
\newblock \url{http://arxiv.org/abs/1010.2481}.

\bibitem[Gomes and Koslowski, 2012]{Gomes:2012aa}
Gomes, H. and Koslowski, T. (2012).
\newblock Coupling shape dynamics to matter gives spacetime.
\newblock \url{http://arxiv.org/abs/1110.3837v2}.

\bibitem[Gomes and Koslowski, 2013]{Gomes:2013aa}
Gomes, H. and Koslowski, T. (2013).
\newblock Frequently asked questions about shape dynamics.
\newblock {\em Foundations of Physics}, 43(12):1428--1458.
\newblock \url{http://arxiv.org/abs/1211.5878v2}.

\bibitem[Gryb and Th{\'e}bault, 2016]{Gryb:2015aa}
Gryb, S. and Th{\'e}bault, K. P.~Y. (2016).
\newblock Time remains.
\newblock {\em British Journal for the Philosophy of Science}.
\newblock DOI 10.1093/bjps/axv009.

\bibitem[Hall, 2009]{Hall:2009aa}
Hall, N. (2009).
\newblock Humean reductionism about laws of nature.
\newblock Unpublished manuscript. http://philpapers.org/rec/halhra.

\bibitem[Huggett, 2006]{Huggett:2006aa}
Huggett, N. (2006).
\newblock The regularity account of relational spacetime.
\newblock {\em Mind}, 115(457):41--73.

\bibitem[Malament, 1977]{Malament:1977aa}
Malament, D.~B. (1977).
\newblock The class of continuous timelike curves determines the topology of
  spacetime.
\newblock {\em Journal of Mathematical Physics}, 18(7):1399--1404.

\bibitem[Miller, 2014]{Miller:2014aa}
Miller, E. (2014).
\newblock Quantum entanglement, {B}ohmian mechanics, and {H}umean
  supervenience.
\newblock {\em Australasian Journal of Philosophy}, 92:567--583.

\bibitem[Pooley, 2001]{Pooley:2001aa}
Pooley, O. (2001).
\newblock Relationalism rehabilitated? {II}: {R}elativity.
\newblock http://philsci-archive.pitt.edu/221/.

\bibitem[Pooley, 2013]{Pooley:2013aa}
Pooley, O. (2013).
\newblock Substantivalist and relationalist approaches to spacetime.
\newblock In Batterman, R., editor, {\em The {O}xford handbook of philosophy of
  physics}, pages 522--586. Oxford: Oxford University Press.

\bibitem[Rovelli, 1997]{Rovelli:1997aa}
Rovelli, C. (1997).
\newblock Halfway through the woods: contemporary research on space and time.
\newblock In Earman, J. and Norton, J., editors, {\em The cosmos of science},
  pages 180--223. Pittsburgh: University of Pittsburgh Press.

\bibitem[Weyl, 1949]{Weyl:1949aa}
Weyl, H. (1949).
\newblock {\em Philosophy of mathematics and natural science}.
\newblock Princeton: Princeton University Press.

\end{thebibliography}

\end{document}